\documentclass[nofootinbib,twocolumn,aps,prd,amsmath,superscriptaddress]{revtex4-1}
\usepackage{setspace}
\usepackage{color}
\usepackage{fancyhdr}
\usepackage{graphicx}
\usepackage[ansinew]{inputenc}
\usepackage{amssymb}
\usepackage{amsmath}
\usepackage{cancel}
\usepackage{float}
\usepackage{subfigure}
\usepackage{cancel}
\usepackage{tabularx}
\begin{document}
\title{Confirming the existence of twin stars in a NICER way}

\author{Jan-Erik Christian}
\email{christian@astro.uni-frankfurt.de}
\affiliation{Institut f\"ur Theoretische Physik, Goethe Universit\"at Frankfurt, Max von Laue Strasse 1, D-60438 Frankfurt, Germany}
\author{J\"urgen Schaffner-Bielich}
\email{schaffner@astro.uni-frankfurt.de}
\affiliation{Institut f\"ur Theoretische Physik, Goethe Universit\"at Frankfurt, Max von Laue Strasse 1, D-60438 Frankfurt, Germany}
\begin{abstract}
We demonstrate that future radius measurement of the NICER mission have the potential to reveal the existence of a strong phase transition in dense neutron star matter by confirming the existence of so called twin stars, compact star configurations with the same mass but different radii. The latest radius constraints from NICER for the pulsars J0740+6620 as well as J0030+0451 are discussed using relativistic mean field equations of state with varying
stiffness, connected with a first order phase transition to quark matter. 
We show, that twin star solutions are compatible with the new radius constraint but are located at radii \emph{below} the present constraints from NICER serving as a smoking gun for a strong phase transition in neutron star matter. This scenario is realized if a strong phase transition takes place in neutron stars of the first branch with masses above $2\,M_\odot$.
\end{abstract}
\maketitle
\section{Introduction}\label{incanus}
The observation of neutron stars has provided one of the most successful possibilities of constraining the equation of state (EoS) of dense strongly interacting matter in the past few years.\\ 
Due to the strong correlation between the equation of state and the mass-radius relation of neutron stars the most intuitive method to constrain the EoS is determining mass and radius of known neutron stars as precise as possible. Currently the most reliable radius measurements are provided by NICER for the pulsars J0030+0451  \cite{Miller:2019cac,Riley:2019yda,Raaijmakers:2019qny} and more recently J0740+6620  \cite{Miller:2021qha,Riley:2021pdl,Raaijmakers:2021uju}. It has been speculated that the new measurement points to a stiffer EoS featuring larger neutron star radii \cite{Pang:2021jta,Annala:2021gom,Huth:2021bsp}. 
The fact, that multiple neutron stars with masses above $2.0\,M_\odot$ are known \cite{Demorest:2010bx,Antoniadis:2013pzd,Fonseca:2016tux,Cromartie:2019kug,Nieder:2020yqy} suggest a stiff EoS as well, since soft EoSs are known to generate less massive maximum mass configurations.\\
Gravitational wave detectors like LIGO open up the possibility to use the tidal deformability of neutron stars as an additional constraint on the EoS. The gravitational event GW170817 points to soft EoSs that feature more compact neutron stars \cite{Paschalidis:2017qmb,Alvarez-Castillo:2018pve,Christian:2018jyd,Montana:2018bkb,Sieniawska:2018zzj,Christian:2019qer}.\\
Many models for the EoS are purely hadronic in nature. However, the high pressure at the center of a neutron star suggests, that a deconfined quark core might be present. Such a phenomenon is called a hybrid star \cite{Ivanenko:1965dg,Itoh:1970uw,Alford:2004pf,Coelho:2010fv,Chen:2011my,Masuda:2012kf,Yasutake:2014oxa,Zacchi:2015oma}. The phase transition can cause a gap in the mass-radius relation, in which case a second branch will form. This second branch necessarily contains a hybrid star with the same mass but a smaller radius than at least one neutron star from the original branch. These two stars are referred to as twin stars \cite{Kampfer:1981yr,Glendenning:1998ag,Schertler:2000xq,SchaffnerBielich:2002ki,Zdunik:2012dj,Alford:2015dpa,Blaschke:2015uva,Zacchi:2016tjw,Alford:2017qgh,Christian:2017jni,Blaschke:2019tbh,Jakobus:2020nxw}.\\
In the following we will apply a relativistic mean field equation of state \cite{PhysRev.98.783,Duerr56,Walecka74,Boguta:1977xi,Serot:1984ey,Mueller:1996pm,Typel:2009sy,Hornick:2018kfi} as the hadronic EoS. Such an EoS can be parameterized \cite{ToddRutel:2005fa,Chen:2014sca,Hornick:2018kfi}, which allows us to adjust its stiffness via the effective nucleon mass, where high effective masses lead to softer equations of state.\\
It has been shown, that an effective mass of at least $m^{*}/m\ge0.60$ is necessary for an EoS to be compact enough to be compatible with GW170817 \cite{Hornick:2018kfi} unless a phase transition at masses below $2\,M_\odot$ is present \cite{Christian:2019qer}.\\
We find that a phase transition has to take place at such masses in order to fit the new NICER constraint from J0740+6620. However, as we will show the possible existence of twin stars is not ruled out by the recent NICER observations. On the contrary, we will demonstrate that the planned future observations of NICER \cite{NICERweb1} have the potential to unequivocally confirm the existence of twin stars. This possibility would imply a significantly smaller minimal radius for neutron stars than purely hadronic EoSs allow for.\\

\section{Theoretical Framework}
\subsection{Phase Transition}
We consider a first order phase transition at high baryonic densities from hadronic to quark matter EoS. We employ the well known constant speed of sound approach \cite{Zdunik:2012dj,Alford:2014dva,Alford:2015gna}:
\begin{equation}
	\epsilon(p) =
	\begin{cases} 
		\epsilon_{HM}(p)	&  p < p_{trans}\\
		\epsilon_{HM}(p_{trans})+\Delta\epsilon + c_{QM}^{-2}(p-p_{trans})	& p > p_{trans}\\
	\end{cases}
\end{equation} 
where $\epsilon_{HM}(p_{trans})$ is the energy density at the point of transition and $p_{trans}$  the transitional pressure.
The discontinuity in energy density at the transition is $\Delta\epsilon$. 
In order to achieve the stiffest possible EoSs and thus the greatest possible range of mass-radius relations we set $c_{QM}=1$ using natural units. Lower values of $c_{QM} $ can not generate higher maximal masses, if the other parameters are identical and $c_{QM}=1$ allows for the greatest range of parameters.
\subsection{Twin Stars}\label{Tevildo}
A first order phase transition can give rise to the phenomenon of "twin stars", which are compact stars with identical masses, but different radii  \cite{Kampfer:1981yr,Glendenning:1998ag,Schertler:2000xq,SchaffnerBielich:2002ki,Zdunik:2012dj,Alford:2015dpa,Blaschke:2015uva,Zacchi:2016tjw,Christian:2017jni}. When analyzing such EoSs it can be useful to classify the twin star solutions into four distinct categories, as shown in \cite{Christian:2017jni}. In a twin star mass-radius relation the maximum of the hadronic branch is referred to as the first maximum $M_1$ and the maximum of the hybrid branch as the second maximum $M_2$. The mass value of the first and second maximum are determined by the values of $p_{trans}$ and $\Delta\epsilon$  \cite{Christian:2017jni}. This behavior is explained by the observation, that the shape of the second branch is correlated with the value of $p_{trans}$, while its position is strongly influenced by the value of $\Delta\epsilon$. High values of $p_{trans}$ lead to high masses in the first maximum and flat second branches. Low values of $\Delta\epsilon$ lead to a second branch near the discontinuity (i.e. a high mass at the second maximum).
Based on this observation the categories can be defined as:
\begin{itemize}
	\item [\textbf{I:}] Both maxima exceed $2\,M_\odot$, which implies high values of $p_{trans}$ and a nearly flat second branch. As a result the maximal mass of a category I case is usually a pure hadronic neutron star.
	\item [\textbf{II:}] Only the first maximum reaches $2\,M_\odot$, which again requires a high value of $p_{trans}$. The flat second branch is moved to lower values of mass and radius.
	\item [\textbf{III:}]The first maximum is in the range of $2\,M_\odot \geq M_{max_1} \geq 1\,M_\odot$, while the second maximum exceeds $2\,M_\odot$. Accordingly, the transitional pressure is lower than in the previous categories and the second branch becomes steeper. The most massive star in these configurations is always a hybrid star.
	\item [\textbf{IV:}] Like category III  the second maximum exceeds $2\,M_\odot$, however the first maximum is below even $1\,M_\odot$. The second branch is at its steepest slope here.
\end{itemize}
Due to the quark matter dominated EoSs in category IV the hadronic part can be nearly arbitrarily soft and the combination can still reach $2\,M_\odot$. Category I-III do not allow for extremely soft nuclear EOSs with effective masses of $m^{*}/m\ge0.75$ \cite{Christian:2019qer}. For low values of $\Delta\epsilon$ the minimum gets closer to the first maximum until no mass gap is present and the phase transition is only noticeable by a kink in the mass-radius relation. In the following we demand a mass gap of $0.1\,M_\odot$ when talking about twin stars, smaller mass-gaps will be considered hybrid star configurations.  We find, that a $\Delta\epsilon = 350\,\mathrm{MeV/fm^3}$ jump in energy density always results in at least a $0.1\,M_\odot$ mass gap \cite{Christian:2020xwz}. However smaller values of $\Delta\epsilon$ can suffice as well.\\

\subsection{Hadronic Equation of State} 
As hadronic equation of state we use a generalized relativistic mean field approach, originally introduced by Todd-Rudel et al. \cite{ToddRutel:2005fa} (see also: Chen et al. \cite{Chen:2014sca}). The  main advantage of this approach is that the slope parameter $L$, the symmetry energy $J$ and the effective nucleon mass $m^*/m$ are adjustable. 
Previous works have shown, that for this particular EoS the choices of $L$ and $J$ have no significant impact on the mass-radius relation \cite{Hornick:2018kfi}. We fix the values $L=60\,\mathrm{MeV}$ and $J=32\,\mathrm{MeV}$ in accordance with our previous publications \cite{Christian:2019qer,Christian:2020xwz}. Note that the stiffness of an EoS relates to the value of $m^{*}/m$  \cite{1983PhLB..120..289B}. Lower values of the effective mass parameter result in a stiffer EoS. Since soft EoSs are favored by GW170817, this means that effective masses $m^{*}/m<0.65$ are not compatible with GW170817 \cite{Hornick:2018kfi} unless a category III or IV phase transition is present \cite{Christian:2019qer}.

\section{Overview of possible phase transitions}
\begin{figure}
	\centering				
	\includegraphics[width=8.6cm]{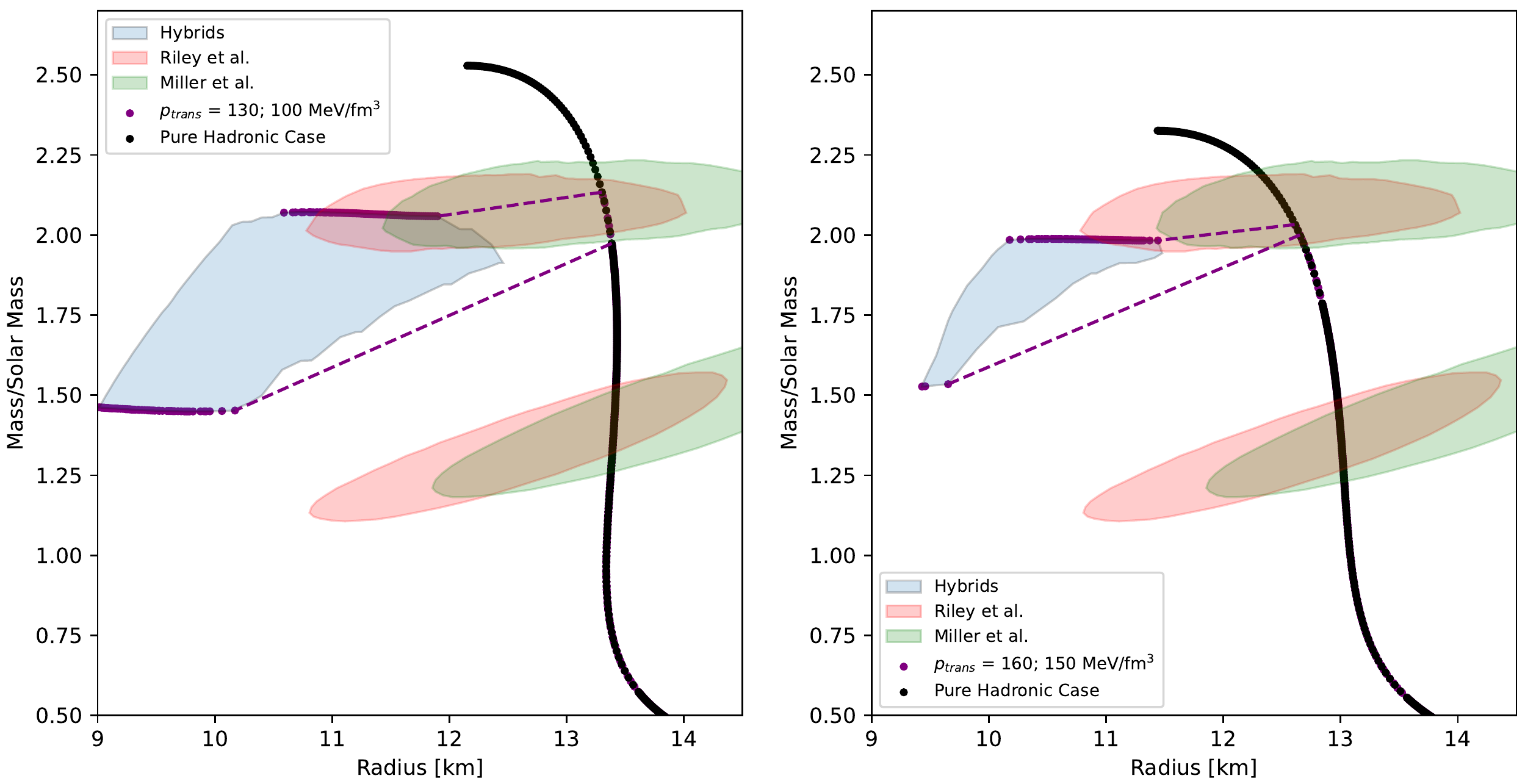}
	\caption{\footnotesize The areas in which the second branches of category I and II phase transition can be located  for EoSs with an effective mass of $m^*/m = 0.65$ (right) and $m^*/m = 0.70$ (left) are shaded blue. The red ellipses indicate the $1\sigma$ ellipses for J0740+6620 (upper ellipse) and J0030+0451 (lower ellipse) as determined by Riley et al. \cite{Riley:2019yda,Riley:2021pdl}. The green ellipses indicate the $1\sigma$ error ellipses for J0740+6620 (upper ellipse) and J0030+0451 (lower ellipse) as determined by Miller et al. \cite{Miller:2019cac,Miller:2021qha}. The purple lines follow the EoSs with the highest and lowest maximal masses in the second branch. Their transitional pressures are stated in units of $\mathrm{MeV/fm^3}$. This demonstrates, that a single EoS can contain neutron stars with radii $<11\,\mathrm{km}$ as well as $M > 2\,M_\odot$. The black line represents the hadronic case without a transition.}
	\label{Region}
\end{figure}
\begin{figure}
	\centering				
	\includegraphics[width=8.6cm]{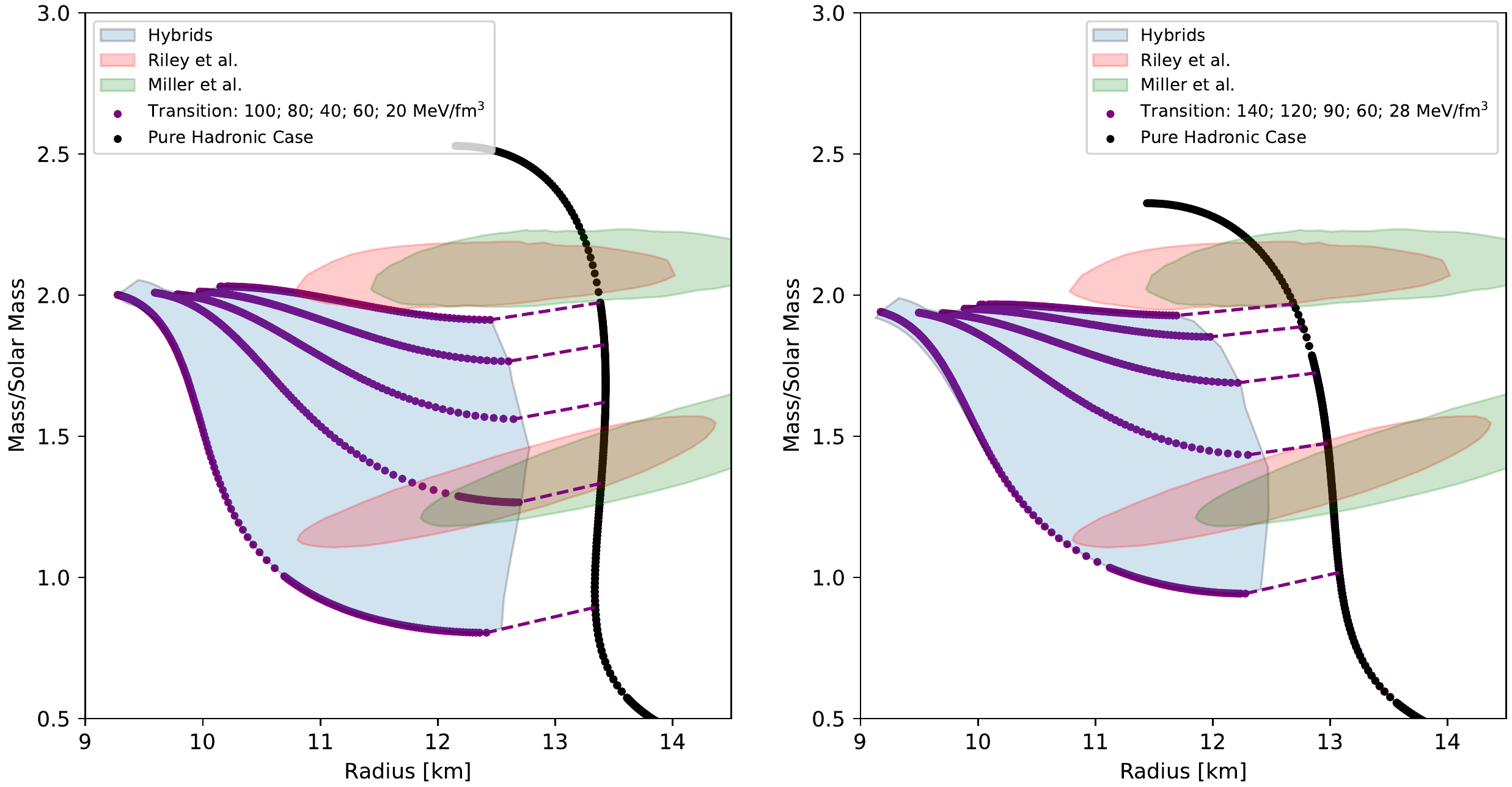}
	\caption{\footnotesize All category III phase transitions with an effective mass of $m^*/m = 0.65$ (right) and $m^*/m = 0.70$ (left) compatible with GW170817 \cite{Abbott:2018wiz} and the $2\,M_\odot$ mass constraint. The purple lines show a few representative examples of category III phase transitions. Their transitional pressures are stated in the key. The lowest second branch line is generated by a jump of 380 $\mathrm{MeV/fm^3}$, all other purple examples have a jump of 350 $\mathrm{MeV/fm^3}$. We can see, that category III can not fulfill the new NICER constraints \cite{Miller:2019cac,Miller:2021qha}, even though small jumps in energy density can fit the previous NICER constraint and the $2\,M_\odot$ constraint quite well.}
	\label{RegionCIII}
\end{figure}
\begin{figure}
	\centering				
	\includegraphics[width=8.6cm]{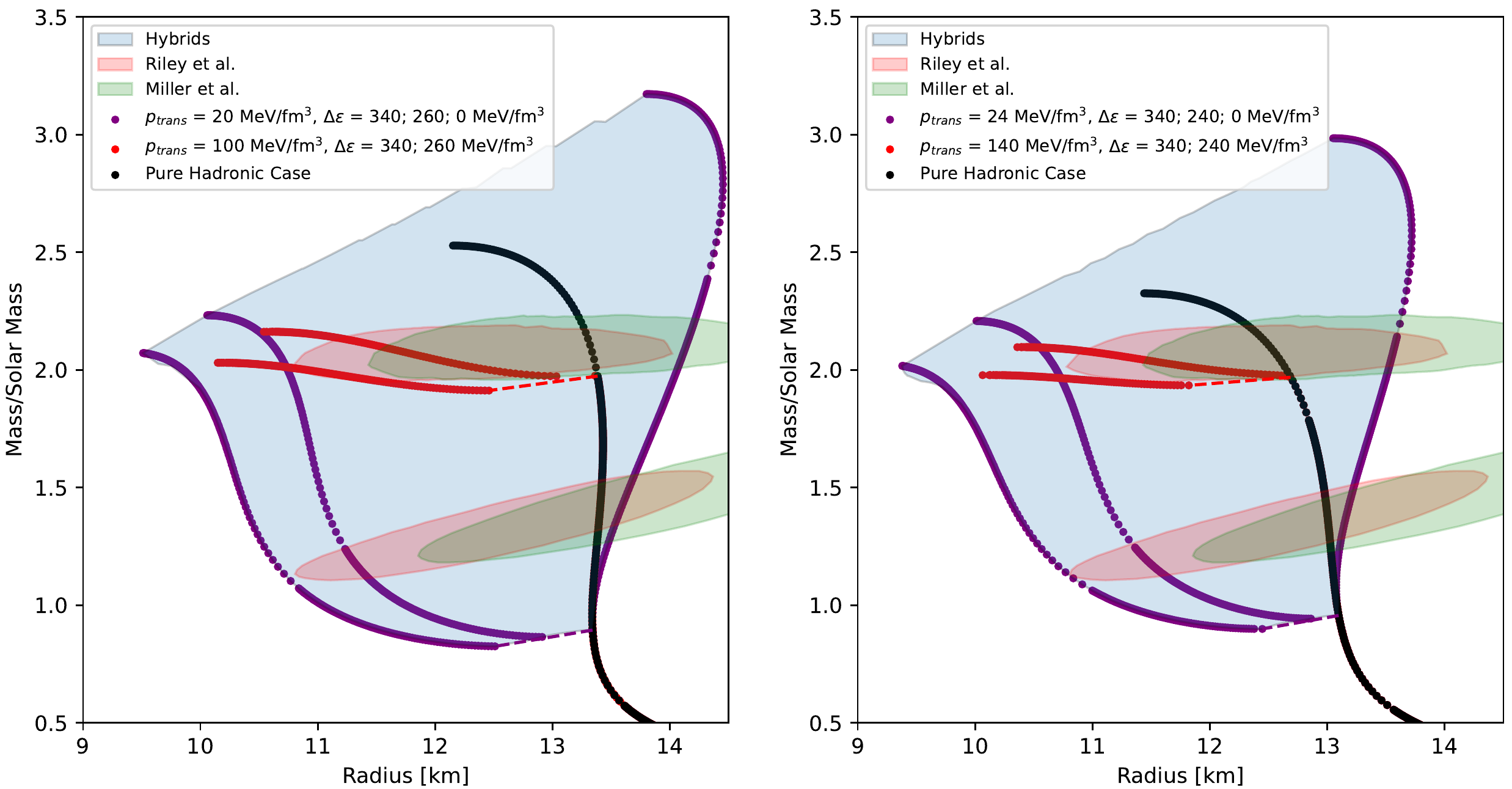}
	\caption{\footnotesize Hybrid star configurations cover a large area of possible mass-radius relations. The purple lines show three examples for the earliest phase transition at a mass of $1\,M_\odot$. The examples where chosen, because they include the limiting cases ($\Delta\epsilon = 340\,\mathrm{MeV/fm^3}$ and $\Delta\epsilon = 0\,\mathrm{MeV/fm^3}$), as well as the smallest jump in energy density not compatible with the new NICER data for this transitional pressure. The red lines represent a transition at $2\,M_\odot$ with $\Delta\epsilon=340\,\mathrm{MeV/fm^3}$ and a jump equal to the one for the purple line in the middle.}
	\label{RegionHybrid}
\end{figure}
In this section we provide an overview of constraint put on twin star solutions by the measurement of J0740+6620.
We only need to consider effective masses of $m^*/m=0.65$ and $m^*/m=0.70$ because effective masses $m^*/m < 0.65$ cannot generate $2\,M_\odot$ in their first branch and still be compatible with GW170817 \cite{Christian:2019qer}. If the $2\,M_\odot$ constraint can only be satisfied in the second branch the radii of this mass configuration lie outside of the new NICER measurement, which we will demonstrate when discussing Fig. \ref{RegionCIII}.\\
In Fig. \ref{Region} category I and II twin star solutions are depicted for the effective masses $m^*/m=0.65$ (left) and $m^*/m=0.70$ (right). The area in which the mass-radius relations can be located is shaded blue and the limiting cases are shown by purple lines generated from $\Delta\epsilon=350\,\mathrm{MeV/fm^3}$ and $\Delta\epsilon=1000\,\mathrm{MeV/fm^3}$ jumps in energy density. The transitionless case is added in black.
The values of $p_{trans}$ are given in the key of Fig. \ref{Region}. The red and green ellipses are the $1\sigma$ error ellipses for J0740+6620 (upper ellipse) and J0030+0451 (lower ellipse) taken from Miller et al. Riley et al. \cite{Riley:2019yda,Riley:2021pdl} (red) and \cite{Miller:2019cac,Miller:2021qha} (green). Fig. \ref{Region} demonstrates, that the new NICER measurement of J0740+6620 \cite{Miller:2021qha,Riley:2021pdl,Raaijmakers:2021uju} does not necessitate a radius constraint for all neutron stars of about $R\ge11\,$km, as some groups suggest \cite{Pang:2021jta,Annala:2021gom,Huth:2021bsp}. If twin star solutions exist there exists a second branch at lower radii while being compatible with the radius constraints from NICER with the first branch. The second branch can be located at masses and radii as low as about $1.46\,M_\odot$ and $9\,\mathrm{km}$ for the case $m^*/m=0.65$ and around $1.55\,M_\odot$ at $9.6\,\mathrm{km}$ for $m^*/m=0.70$. Therefore, measurement of a hybrid star from the second branch of a category I or II case could confirm the existence of twin stars.\\
In Fig. \ref{RegionCIII} we demonstrate, that a twin star phase transition can not be achieved in category III. It shows all category III configurations in agreement with the previous NICER, mass and GW170917 constraints. The limiting cases and a few exemplary cases are marked in purple, their transitional pressures are listed in order from top to bottom in the key. All cases feature a jump of $\Delta\epsilon = 350\,\mathrm{MeV/fm^3}$, with the exception of the earliest phase transition, which features a jump of $\Delta\epsilon = 380\,\mathrm{MeV/fm^3}$ in order to demonstrate the lower limit of the area in which second branches are possible. It is necessary for category III to feature jumps in energy density of $\Delta\epsilon \le 380\,\mathrm{MeV/fm^3}$, otherwise the $2\,M_\odot$ constraint can not be fulfilled.
It becomes clear, that the new NICER constraint can not be covered by the category III configurations. Only high transitional pressures come close, however at those values of $p_{trans}$ the configurations are virtually indistinguishable from early category I phase transitions. Lower effective masses behave similarly, but they also have to cope with the problem, that higher transitional pressures make them incompatible with GW170817 \cite{Christian:2019qer}, i.e. the radius gets too large for a $1.4\,M_\odot$ neutron star.\\
When considering phase transitions with discontinuities at $\Delta\epsilon < 350\,\mathrm{MeV/fm^3}$ the new measurement of NICER is easily reachable, which is shown in Fig. \ref{RegionHybrid}. The reason for this is, that hybrid star configurations cover a large area of possible mass radius relations. However this also means, that they are difficult to distinguish from ordinary neutron stars without a phase transition.
The purple lines indicate the limiting cases with a transition at $1\,M_\odot$. Their respective jumps in energy density are stated in the key. The purple line in the middle shows the smallest possible $\Delta\epsilon$ outside of the new NICER data. The red lines represent a transition at $2\,M_\odot$ with $\Delta\epsilon=340\,\mathrm{MeV/fm^3}$ and a jump equal to the one for the purple line in the middle.\\
At higher transitional pressure this small discontinuity is not enough to generate a second branch, instead the mass-radius relation has a kink at the point of transition, which is demonstrated with the upper red line. This line has the same jump in energy density, that previously was unable to reach the ellipsis, but at a transition point of $2\,M_\odot$. The absence of a mass gap means such phase transitions would not be detectable via observation of "rising twins", where the more massive twin also has a greater radius \cite{Schertler:2000xq}. Furthermore the difference in radius of similarly massive stars becomes smaller. This problem intensifies for smaller jumps in energy density, which makes hybrid star configurations hard to distinguish from pure hadronic mass-radius relations. The lower red lines show phase transitions at $2\,M_\odot$ with a jump of $\Delta\epsilon = 340\,\mathrm{MeV/fm^3}$, which feature rising twins. However these configurations are outside of the new NICER constraint as well as nearly indistinguishable from category I configurations.

\section{Discussion}\label{conclusion}
We have shown that twin stars are not ruled out by the newest NICER measurement. However, twin stars can only be generated if the new constraint is fulfilled before the phase transition takes place. This allows for hybrid stars with significantly smaller radii than the NICER constraint in the second branch.
We find that future NICER measurements could confirm the existence of twin stars, by finding neutron stars from these second branches. Pulsars with known masses below $2\,M_\odot$, like J1614-2230 and J0437-4715 are of particular interest. Finding that one of those two soon to be measured neutron stars \cite{NICERweb1} has a smaller radius than the newest measurement of J0740+6620 would confirm phase transitions, specifically ones with a large jump in energy density.
The measurement of J0740+6620 does suggest a phase transition at high masses in the first branch. This inevitably leads to a flat second branch, with a  great range of possible radii at a specific mass. This mass depends on the discontinuity of energy density in the phase transition. The measurement of a neutron star with a small radius and known mass would therefore not only point strongly to the existence of twin stars, but would also constrain the parameters of the required phase transition well.
Finally we find, that phase transitions with parameters that do not support twin stars fit the data as well, but are unlikely to be detectable by NICER. Like twin star cases the phase transitions without twin stars favor a high transitional pressure. \\
During completion of this work a related work was published by Li et al. \cite{Li:2021sxb}. They find that twin stars are supported by the new NICER data. However, in their work the second branch always contains the most massive neutron star and is therefore within the newest NICER data. We show, that an EoS can fulfill all constrains and feature a second branch outside of the current data, which would be a stronger indication of a phase transition within neutron stars.
\begin{acknowledgments}
 J.E.C. thanks the Giersch foundation for their support with a Carlo and Karin Giersch Scholarship. 
\end{acknowledgments}

\bibliographystyle{apsrev4-1}
\bibliography{neue_bib_JSB}
\end{document}